# An Adaptive Power Efficient Packet Scheduling Algorithm for Wimax Networks

R Murali Prasad
Department of Electronics and Communications
MLR Institute of Technology, Hyderabad
muraliprasadphd@gmail.com

Dr.P. Satish Kumar
Professor, Department of Electronics and Communications
CVR College of Engineering, Hyderabad
satishkumar_1968@redifmail.com

*Abstract*—Admission control schemes and scheduling algorithms are designed to offer QoS services in 802.16/802.16e networks and a number of studies have investigated these issues. But the channel condition and priority of traffic classes are very rarely considered in the existing scheduling algorithms. Although a number of energy saving mechanisms have been proposed for the IEEE 802.16e, to minimize the power consumption of IEEE 802.16e mobile stations with multiple real-time connections has not yet been investigated. Moreover, they mainly consider non real- time connections in IEEE 802.16e networks. In this paper, we propose to design an adaptive power efficient packet scheduling algorithm that provides a minimum fair allocation of the channel bandwidth for each packet flow and additionally minimizes the power consumption. In the adaptive scheduling algorithm, packets are transmitted as per allotted slots from different priority of traffic classes adaptively, depending on the channel condition. Suppose if the buffer size of the high priority traffic queues with bad channel condition exceeds a threshold, then the priority of those flows will be increased by adjusting the sleep duty cycle of existing low priority traffic, to prevent the starvation. By simulation results, we show that our proposed scheduler achieves better channel utilization while minimizing the delay and power consumption.

*Keywords- QOS; Packet Scheduling Algorithm; IEEE 802.16/802.16e; WiMAX Networks; delay and power consumption.*

## I. INTRODUCTION

### A. WiMAX Networks

WiMAX (Worldwide interoperability for Microwave access) or IEEE 802.16 is regarded as a standard for metropolitan area networks (MANs) [2]. It is one among the most reliable wireless access technologies for upcoming generation all-IP networks. In reality, this access technology enables obtaining high bit rate and reaching large areas with a single Base Station (BS), and because of this it provides to operators the option of supplying connectivity to end users in an economical way [3]. It is a reliable choice to offer last-mile access in wireless metropolitan area network (WMAN) together with the merits of low cost, high speed, rapid and easy deployment, such that a large number of applications can be applied also in the areas where the installation of wired infrastructure is cost-effective or technically achievable [4]. In consequence to the characteristics of WiMax, it can be widely employed in several related fields, comprising of mobile service, mobile commerce, mobile entertainment, mobile learning and mobile healthcare [5].

Fixed subscriber stations (SSs) and mobile subscriber stations (MSSs) remain in contact with BSs by means of air interfaces [2]. Even though the deployment and the utilization of this standard have begun, the exploitation of WiMAX networks is still restricted to certain situations. Research works on WiMAX access networks is still taking place, because several topics have yet to be described to permit and optimize the utilization of this technology in upcoming generation networks [3].

Instant progress of wireless technology, together with the development of the internet, has augmented the demand for wireless data services. Next-generation wireless communication systems are anticipated to offer an extensive range of services with excessive and time-varying data rate requirements, with several and variable quality of service (QoS) constraints. Traffic on 4G networks namely WiMAX is heterogeneous with random mix of real and non-real time traffic with applications needing widely varying and miscellaneous QoS guarantee [7, 8].

The 802.16 standard provides two modes for sharing the wireless medium:

- Point-to-Multipoint (PMP) and
- Mesh (optional).

In the PMP mode, the nodes are arranged to form a cellular-like structure, where a base station (BS) aids a set of subscriber stations (SSs) within the same antenna sector in a broadcast mode, with all SSs obtaining the same transmission from the BS. Transmissions from SSs are targeted to and synchronized by the BS. On the other hand, in Mesh mode, the nodes are organized ad hoc and scheduling is distributed among them. In the IEEE 802.16 standard, uplink (from SS to BS) and downlink (from BS to SS) data transmissions are frame-based [1].

### B. Scheduling Issues in WiMAX

Scheduling is the process of allocating time slots to SSs in each frame so that the transmissions of nearby SSs will not cause collision, and global fairness among SSs can be maintained [8].

WiMAX mesh mode employs two scheduling methods for assigning network resources and managing network access: Centralized and Distributed. When employing centralized scheduling, the BS collects the requests of various SSs that are linked to it developing a mesh tree and assigns resources locally. However, in distributed scheduling, requests and grants are transferred in the extended neighborhood that includes the neighbors and their direct neighbors. The standard offers signaling control messages for centralized and distributed scheduling also but avoids the scheduling algorithms open for the vendors. Besides, it characterizes the control and management messages to be employed for constructing the mesh, without describing the routing metrics to be employed [9].

Several scheduling algorithms were proposed, such as adaptive uplink and downlink bandwidth adjustment, to achieve higher transmission performance. Even with these methods and algorithms, the packet queuing delay and radio link utilization of an IEEE 802.16-based network cannot be greatly improved due to its frame structure and bandwidth requesting/granting procedure [2].

Numerous studies examined the power consumption problems of IEEE 802.16e and recommended algorithms to establish the sleep interval in augmenting its energy efficiency. But, the entire study mainly takes into account non real- time connections in IEEE 802.16e networks. Noticeably, devoid of a proper schedule of the sleep-mode operations for multiple real-time connections on a mobile station, the power consumption of a mobile station cannot be lowered even when the sleep mode is applied [10].

In this paper, we propose to design an ideal adaptive power efficient packet scheduling algorithm that provides a minimum fair allocation of the channel bandwidth for each packet flow and additionally minimizes the power consumption.

## II. RELATED WORK

Lien-Wu Chen et al [2] have studied how to exploit spectral reuse in resource allocation in an IEEE 802.16 mesh network, which includes routing tree construction (RTC), bandwidth allocation, time-slot assignment, and bandwidth guarantee of real-time flows. Their proposed spectral reuse framework covers bandwidth allocation at the application layer, RTC and resource sharing at the medium access control (MAC) layer, and channel reuse at the physical layer. Also their paper formally quantifies spectral reuse in IEEE 802.16 mesh networks and exploits spectral efficiency under an integrated framework.

Hanwu Wang et al [4] have studied the radio resource scheduling and traffic management in WiMAX networks. Their proposed scheme is scalable in the sense that different types of sessions can be integrated into a unified scheduling process to satisfy a flexible QoS (Quality of Service) requirement. Their scheme is adaptive that all the traffic rates can be controlled and tuned fast under different network traffic-load conditions. Also they have adopted the classical control theory method into their proposed mechanism, which helps to achieve high efficiency (utilization), perfect traffic throughput, fairness, and system stability.

Chakchai So-In et al [6] have defined the Generalized Weighted Fairness (GWF) criterion that allows carriers to implement either of the two fairness criteria applying into a Mobile WiMAX environment. In addition, they show how a scheduling algorithm can use the GWF criterion to achieve a general weighted fair resource allocation in IEEE 802.16e Mobile WiMAX networks.

Shiao-Li Tsao et al [10] have proposed two energy-efficient packet scheduling algorithms for real-time communications in a Mobile WiMAX system. Their schemes not only guarantee the quality of services (QoSs) of real-time connections but also minimize power consumption of mobile stations by maximizing the length of a sleep period in the type-two power-saving class defined in the IEEE 802.16e, without violating QoSs of all connections.

Guowang Miao et al [11] have applied the utility-based framework to evaluate system performance improvements for a downlink 802.16 OFDMA wireless communication system. They consider scheduling across a mix of rate-adaptive as well as real-time service classes. Also they focused on channel-aware scheduling only.

S. Lakani et al [12] have proposed a new approach to improve distributed scheduling efficiency in IEEE 802.16 mesh mode, with respect to network condition in every transferring opportunity. Their proposed approach can reduce transmission delay.

## III. PROPOSED SCHEDULING

### A. System Design

WiMAX system has five types of the traffic service, namely

- UGS (Unsolicited Grant Service),
- rtPS (Real Time Polling Service),
- ertPS (Extended Real Time Polling Service),
- nrtPS (Non- Real Time Polling Service), and
- BE (Best Effort)

The traffic flow is categorized into the following 3 classes:

1. Class1 (UGS,rtPS and ertPS)
2. Class2 (nrtPS)
3. Class2 (BE)

Each node $n_i$ maintains 3 queues $q_{i1}, q_{i2}$ and $q_{i3}$ for the traffic classes Class1, Class2 and Class3 respectively. Each node shares the queuing information with other nodes within the communication range in control frame of the 802.16e. A Channel Condition Estimator (CCE) monitors the channel periodically and estimates the channel state error (SINR).If there is no channel error, then resource are scheduled as per their priority of traffic classes, in a power efficient manner (described in section III.c).

If there is a channel error, then

- Precede the transmission if the node has Class1 packets.
- Otherwise, the transmission is stopped and the allotted slots are assigned to other neighboring nodes.

To minimize power consumption of a mobile station (MS) with multiple real-time connections, we have to determine the length of a sleep period and a listen period under the radio resource and QoS constraints.

Considering a mobile station $j$ with $N$ real-time connections, the QoS parameters of connection $i$ can be denoted as $Q_{ji}\{PS_i, AT_i, D_i\}$, where

- $D_i$ is the delay constraint of any two consecutive packets for connection $i$
- $PS_i$ is the average packet size in bytes for connection $i$
- $AT_i$ is the average inter packet arrival time in milliseconds for connection $i$.

In this paper, these connections could be either downlink from a base station to a mobile station or uplink from a mobile station to a base station.

*B. Channel Error Estimation*

Here, we denote a communication link as $l_i = (s_i, r_i)$, where $s_i$ is the sender and $r_i$ is the receiver node. According to our model, the Packet Reception Rate (PRR) experienced on link $l_i$, in the absence of interference, is given by $f(SNR_i)$, where $SNR_i$ is the signal-to-noise ratio at node $r_i$. Formally, $\text{SNR}_i = \frac{P_i}{N}$ where $P_i$ is the received power at node $r_i$ of the signal transmitted by node $s_i$, and $N$ is the background noise power.

In presence of multiple concurrent transmissions on links $l_1 \cdots l_k$, the PRR on link $l_i = (s_i, r_i)$ is given by $f(SINR_i)$, where $SINR_i$ is the signal-to-noise-and interference ratio measured at $r_i$ when all the $s_j's$ are transmitting. Formally,

$$\text{SNIR}_i = \frac{P_i}{N + \sum_{j \neq i} P_j} \quad (1)$$

where $P_j$ denotes the received power at node $r_i$ of the signal transmitted by node $s_j$, for each $j \neq i$.

*C. Adaptive Sleep Duty Generation*

In a Mobile WiMAX system, a mobile station can switch to sleep mode if there is no packet to send or receive in order to save power. The IEEE 802.16e defines three power-saving classes to accommodate network connections with different characteristics. According to the specification, each connection on a mobile station can be associated with a power-saving class, and connections with a common demand property can be grouped into one power-saving class. The parameters of a power-saving class, i.e. the time to sleep and listen, the length of a sleep period and a listen period can be negotiated by a base station and a mobile station. [10].

If a mobile station establishes multiple connections with different demand properties, the periods that a mobile station can sleep are determined by the sleep-mode behaviors associated with all connections. Obviously, without a proper schedule of the sleep-mode operations for multiple real-time connections on a mobile station, the power consumption of a mobile station might not be reduced even the sleep mode is applied.

Since the periodic power save scheme requires a fixed sleep and listen periods for a station, it might have to stay awake in some frames in the listen period even if there is no packet available. So a non-periodic (NP) scheme, in which length of sleep and listen periods are variable, has to be applied in order to determine the sleep and wakeup cycles in a frame basis. The BS activates this scheme whenever a connection is established or released on a MS to re-schedule the resources in the following frames for the mobile station.

All connections on a mobile station are checked to determine their traffic class and sorted according to their priority level. Within each class, the connections are sorted based on their request dead-lines.

After the scheduler decides the scheduling priorities of connections, the packets from the first priority connection $i$ from the node $j$ are scheduled. Let $RBW_{ij}(k)$ be the request bandwidth of the connection $i$ of the node $j$ in the $k^{th}$ OFDM frame. Let $TBW_j$ be the total available bandwidth in an OFDM frame of duration $T_f$ to the node $j$.

To schedule $RBW_{ij}(k)$, both the bandwidth and delay constraints are to be satisfied. (i.e.)

$$\text{If } RBW_{ij}(k) < TBW_j - ABW_j(m), m \geq k \quad (2)$$

$$If (m - k + 1) \times Tf \leq DC_i \quad (3)$$

Where $ABW_j(m)$ is the already allocated bandwidth for the node $j$ for other connections in the $m^{th}$ frame and $DC_i$ is the delay constraint in milliseconds of any two consecutive packets for connection $i$.

Let $F\{RBW_{ij}\}$ be the set of feasible scheduling frames for $RBW_{ij}$.

To assign the priority and select a frame $F_i \in F\{RBW_{ij}\}$, the following steps are followed.

(i) If $F_i$ is an in-used frame and if the resources of the in-used frames are still available to accommodate $RBW_{ij}$, high priority is assigned for $F_i$.

(ii) If there are two in-used frames $F_i$ and $F_j$, then the priority of assigned for $\min(F_i, F_j)$

(iii) If $F_i, i = 1,2\cdots,n$ are un-used frames, then priority is assigned for the last un-used frame $F_n$. This is because, if a latter frame can be selected, it gains more opportunities to serve other packets in the following OFDM frames

After the above steps, $RBW_{ij}$ is scheduled to the selected frame.

*D. Adaptive Power Efficient Packet Scheduling Algorithm*

1. Get the traffic request from the node $MSS_i$.
2. Estimate the channel error ($SINR$) for $MSS_i$ by (1).
3. If $SNIR < SNIR_{thr}$, where $SNIR_{thr}$ is the threshold value for $SNIR$, then
    3.1 $MSS$ belongs to Group1
 Else
    3.2 $MSS$ belongs to Group2.
 End if.
4. If $MSS$ belongs to Group1,
    4.1 Check the traffic class.
    4.2 Assign priorities as per the traffic class.
    4.3 Allocate the required bandwidth and slots for
    $MSS$ by.
    4.4 Generate sleep duty cycles for the nodes
    satisfying (2) and (3).
 End if
5. If $MSS$ belongs to Group2,
   5.1 Check the traffic class $j$, j = 1 to 3.
   5.1 Add the traffic $TR_k$ into the corresponding queue $q_{ij}$
   5.2 If $j = 1$ and If $Q_{size_{ij}} > Q_{thr\ r}$, where $Q_{size_{ij}}$ is the
    queue length of $q_{ij}$ and $Q_{thr}$ is the maximum queue
    length, then
     5.2.1 Increase the sleep duration for any
     traffic $TR_{ji}$, $j = 2$ or 3 of Group1 node.
     5.2.2 Allocate the slot of $TR_{ji}$ to the excess traffic
     flow $TR_k$
   Else
    Repeat from Step.2
   End if
 End if

## IV. SIMULATION RESULTS

*A. Simulation Model and Parameters*

To simulate the proposed scheme, network simulator (NS2) [13] is used. The proposed scheme has been implemented over IEEE 802.16 MAC protocol. In the simulation, clients (SS) and the base station (BS) are deployed in a 1000 meter x 1000 meter region for 50 seconds simulation time. All nodes have the same transmission range of 250 meters. In the simulation, the video traffic (VBR) and CBR traffic are used.

The simulation settings and parameters are summarized in table I.

TABLE I. SIMULATION SETTINGS

| Area Size | 1000 X 1000 |
|---|---|
| Mac | 802.16 |
| Clients | 2,4,6,8 and 10 |
| Radio Range | 250m |
| Simulation Time | 50 sec |
| Routing Protocol | DSDV |
| Traffic Source | CBR, VBR |
| Video Trace | JurassikH263-256k |
| Physical Layer | OFDM |
| Packet Size | 1500 bytes |
| Frame Duration | 0.005 |
| Rate | 1Mb |
| Error Rate | 0.01,0.02,....0.05 |

*B. Performance Metrics*

We compare our proposed APEPS scheme with the PBS scheme [ ]. We mainly evaluate the performance according to the following metrics:

**Channel Utilization:** It is the ratio of bandwidth received into total available bandwidth for a traffic flow.

**Throughput:** It is the number of packets received successfully

**Average End-to-End Delay:** The end-to-end-delay is averaged over all surviving data packets from the sources to the destinations.

**Average Energy:** It is the average energy consumption of all nodes in sending, receiving and forward operations

The performance results are presented in the next section.

*C. Results*

**A. Effect of Varying MSS**

In our first experiment, the number of MSS is varied as 2, 4, 6, 8 and 10 and we measure the channel utilization, throughput, energy and end-to-end delay.

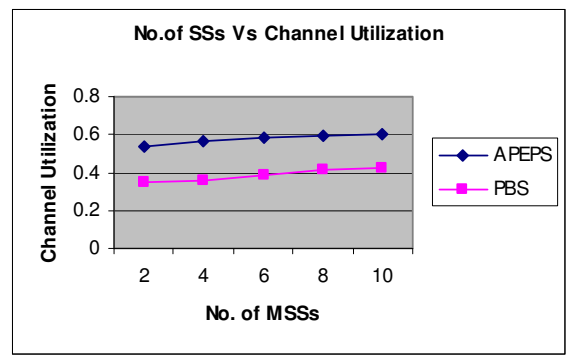

Figure 1. MSS Vs Utilization

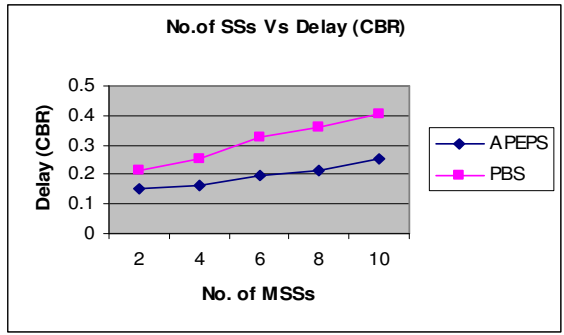

Figure 2. MSS Vs Delay (CBR)

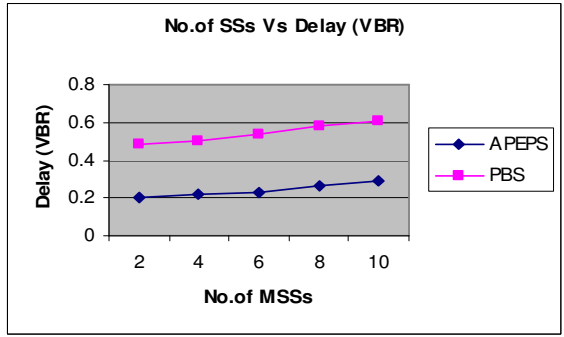

Figure 3. MSS Vs Delay (VBR)

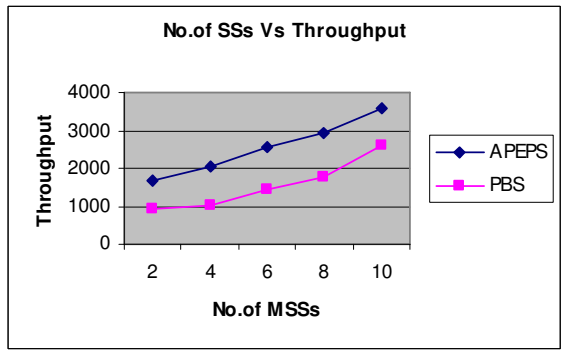

Figure 4. MSS Vs Throughput

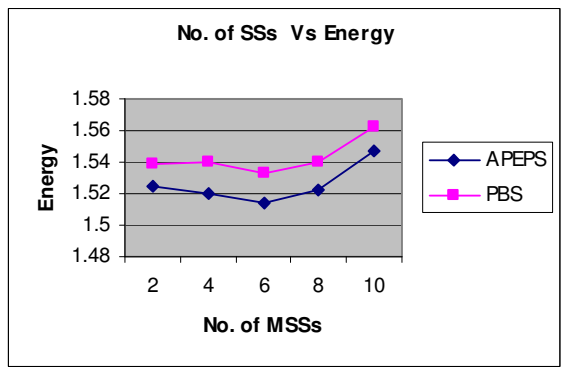

Figure 5. MSS Vs Energy

Figure.1 shows the channel utilization obtained, when the number of MSS is varied. It shows that APEPS has better utilization than the PBS scheme.

Figure.2 and Figure.3 shows the delay of CBR and VBR traffic occurred, when MSS is varied. It shows that the delay of APEPS is significantly less than the PBS scheme for the both the traffics.

Figure.4 shows the throughput obtained with our APEPS scheme compared with PBS scheme. It shows that the throughput of APEPS is more than the PBS, as MSS increases.

The energy consumption of both the schemes is presented in Figure.5. From the figure, we can observe that the energy consumption is less in APEPS when compared to the PBS scheme.

**B. Effect of Varying Channel Error Rates**

In the second experiment, we vary the channel error rate as 0.01, 0.02, 0.03, 0.04 and 0.05. The results are given for class1 and class2 traffics.

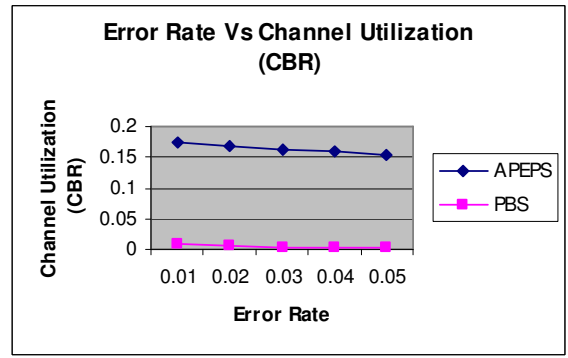

Figure 6. Error Rate Vs Utilization (CBR)

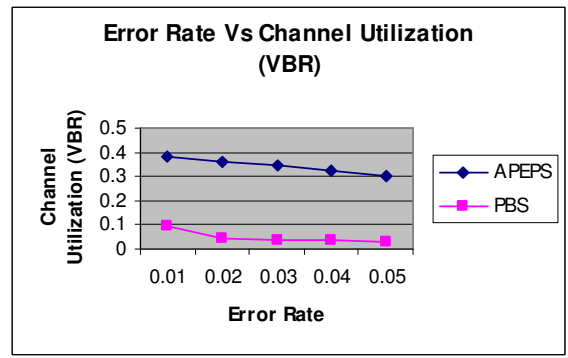

Figure 7. Error Rate Vs Utilization (VBR)

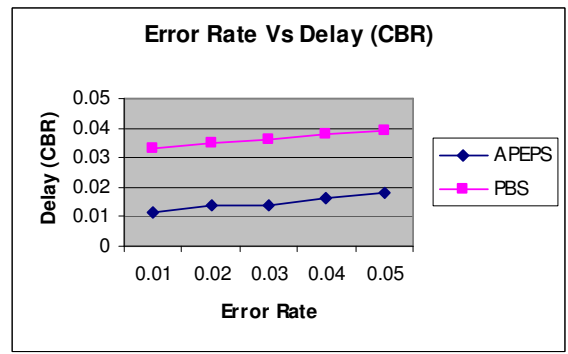

Figure 8. Error Rate Vs Delay (CBR)

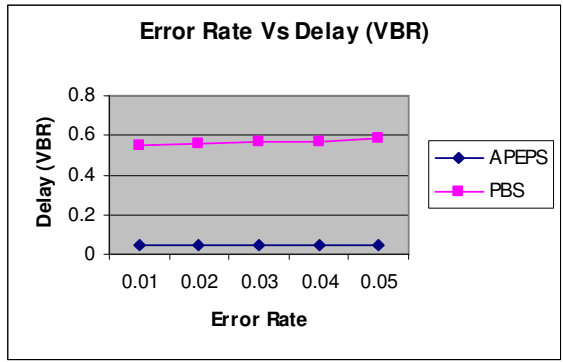

Figure 9. Error Rate Vs Delay (VBR)

Normally, when the channel error rate is increased, the channel utilization of all the flows will tend to decrease. As it can be seen from the figures 6 and 7, the utilization of all the flows slightly decreases, when the error rate is increased. As per the proposed algorithm, the class1 flow is admitted and other types of flows are blocked when there is a channel error. So the utilization for class1 is more, when compared with class2 flows

Figure.6 and Figure.7 shows the channel utilization for CBR and VBR traffics obtained for various error rates. It shows that APEPS has better channel utilization than the PBS scheme.

Figure.8 and Figure.9 shows the delay for the CBR and VBR traffics occurred for various error rates. It shows that the delay of APEPS is significantly less than the PBS scheme.

V. CONCLUSION

In this paper, we have proposed to design an adaptive power efficient packet scheduling algorithm that provides a minimum fair allocation of the channel bandwidth for each packet flow and additionally minimizes the power consumption. In the adaptive scheduling algorithm, packets are transmitted as per allotted slots from different priority of traffic classes adaptively, depending on the channel condition. Based on the channel conditions, the nodes can be classified as Group1 and Group 2. The Group 1 has the MSS with good channel condition whereas the Group 2 has the MSS with bad channel condition. If the MSS belongs to Group1, then the nodes are assigned priority according to the traffic classes and slots are allocated. Group 2 nodes will be in the waiting state in separate queues until their channel condition becomes good. Suppose if the buffer size of the high priority traffic queues with bad channel condition exceeds a threshold, then the priority of those flows will be increased by adjusting the sleep duty cycle of existing low priority traffic, to prevent the starvation. By simulation results, we have shown that our proposed scheduler achieves better channel utilization while minimizing the delay and power consumption.

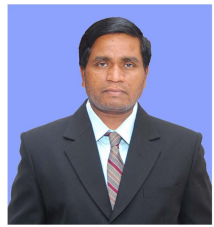

R Murali Prasad received Engineering degree from the Institution of Engineers (i) in 1989 and M.Tech degree from the department of Electronics and Communications, Pondicherry Engineering College in 1993.He worked in various engineering colleges as faculty member. Presently he is working as faculty member in the Department of Electronics and Communications, MLR Institute of technology, Hyderabad. He is pursuing Ph.D at JNT University Anantapur under the guidance of Dr.P, Satish Kumar. His areas of interest are digital communications, control systems and wireless communications.

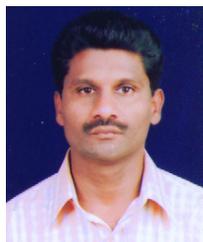

Dr. P. Satish Kumar received B.Tech degree in the Department of Electronics and Communication from Nagarjuna University in 1989 and M.Tech degree from Pondicherry University in 1992. He completed Ph.D degree from JNT University, Hyderabad in the year 2004. He is having 18 years of teaching experience. He has published 15 research papers at national and international level. Presently he is working as professor in the Department of Electronics and Communications, CVR college of engineering, Hyderabad. His research areas are multirate signal processing, image processing and wireless communications.